\documentclass[twocolumn]{aastex63}
\usepackage{inputenc}
\usepackage{comment}
\usepackage{colortbl}
\usepackage{graphicx}
\usepackage{soul}
\usepackage{hyperref}
\usepackage{multirow}
\usepackage{float}

\newcommand\Resolve{{\it XRISM}/Resolve }
\newcommand\FeKa{Fe K$\alpha$ }
\newcommand\FeKone{Fe K$\alpha_1$ }
\newcommand\FeKtwo{Fe K$\alpha_2$ }

\received{12 Dec 2025}
\revised{3 Jan 2026}
\accepted{4 Jan 2026}
\submitjournal{ApJ-L}

\shorttitle{XRISM FeK$\alpha$}
\shortauthors{DiKerby et al. 2025}
\graphicspath{{./}{figures/}}
\begin{document}

\title{Resolving the \FeKa Doublet of the Galactic Center Molecular Cloud G0.11-0.11 with XRISM}

% Steve
\author[0000-0003-2633-2196]{Stephen DiKerby}
\affiliation{Department of Physics and Astronomy \\
 Michigan State University, East Lansing, MI 48820, USA}

% Shuo
\author[0000-0002-2967-790X]{Shuo Zhang}
\affiliation{Department of Physics and Astronomy \\ Michigan State University, East Lansing, MI 48820, USA}

% Kumiko
\author[0000-0002-0726-7862]{Kumiko K. Nobukawa}
\affiliation{Faculty of Science and Enginnering \\ Kindai University, Higashi-osaka, Osaka, Japan 577-8502}

% Masa
\author[0000-0003-1130-5363]{Masayoshi Nobukawa}
\affiliation{Faculty of Education \\ Nara University of Education, Nara, Nara, Japan 630-8502}

% Yuma
\author{Yuma Aoki}
\affiliation{Faculty of Science and Enginnering \\ Kindai University, Higashi-osaka, Osaka, Japan 577-8502}

% Jack
\author[0009-0006-9659-0640]{Jack Uteg}
\affiliation{Department of Physics and Astronomy \\ Michigan State University, East Lansing, MI 48820, USA}

\begin{abstract}

\FeKa line emission from Galactic center molecular clouds can be produced either via fluorescence after illumination by an X-ray source or by cosmic ray ionization. Unparalleled high-resolution X-ray spectroscopy obtained by \Resolve for the galactic center molecular cloud G0.11-0.11 resolves its \FeKa line complex for the first time, and points to a new method for discrimination between the X-ray reflection and cosmic ray ionization models. The \FeKa line complex is resolved into \FeKone at $E_{1} = 6.4040 \: \rm{keV}$ and \FeKtwo at $E_{2}= 6.3910 \:\rm{keV}$. Both lines have non-instrumental FWHM of $\approx 3 \:\rm{eV}$, close to the predicted quantum mechanical width of the lines, suggesting scant other sources of line broadening other than instrumental and quantum effects. We measure a radial velocity of $v_{\rm{LSR}} = 50 \pm 12_{fit} \pm 14_{scale} \:\rm{km/s}$ for G0.11-0.11, achieving the same precision reached by radio observations of such clouds. The high-resolution spectrum tests for the presence of secondary \FeKa lines, expected as a signature of cosmic ray proton/ion ionization. The absence of the secondary lines argues against the cosmic ray ionization model for G0.11-0.11. In the preferred X-ray reflection model, if the illuminating source is Sgr A$^{\star}$, the required luminosity for an X-ray outburst about 200 years ago is $L_8 \approx 10^{38} \:\rm{erg/s}$ in an $8\:\rm{keV}$-wide band at $8\:\rm{keV}$.

\end{abstract}


\section{Introduction and Context}
\label{sec:Intro}

The \FeKa emission line complex near $6.4 \:\rm{keV}$ corresponds to the transition of an electron from the $n=2$ to the $n=1$ levels in neutral ambient iron atoms, and is one of the most potent emission lines studied in X-ray astronomy.  The study of \FeKa line structure has previously been hampered by the limited spectral resolution of most X-ray telescopes. These restrictions have inhibited precise spectroscopic analysis of the component lines that make up the \FeKa complex, \FeKone and \FeKtwo at $6.40401$ and $6.3910 \:~\rm{keV}$ with relative strength $2:1$ \citep{2020PASJ...72L...7O}. The unprecedented spectral resolution of the \Resolve detector with $R = E/\Delta E \approx 1000$ opens up a new regime to probe substructure of the \FeKa complex and other atomic lines.

Several molecular clouds throughout the Galaxy are known to emit copious \FeKa radiation, which has been variously explained using two competing models. In the X-ray reflection scenario, X-ray photons from an external source sweep through the molecular cloud. Photo-ionization and subsequent fluorescence give rise to the observed \FeKa emission, while inverse Compton scattering accounts for an X-ray continuum up to $\sim100 \:\rm{keV}$. \citep{2011ApJ...740..103O,2010ApJ...714..732P,2010ApJ...719..143T}. In the alternative cosmic-ray ionization model, $>\rm{GeV}$ cosmic ray protons, ions, or electrons bombard molecular clouds, producing \FeKa emission via collisional ionization as well as a continuum component via non-thermal Bremsstrahlung.\citep{2012A&A...546A..88T,2009PASJ...61..901D,2011A&A...530A..38C}. 

Although both mechanisms can simultaneously contribute to X-ray emission from Galactic center molecular clouds (GCMC), the X-ray variability of GCMC on the timescale of $\sim10$ years \citep{2013A&A...558A..32C,2025A&A...695A..52S,2015ApJ...815..132Z,2022icrc.confE.288R,2012int..workE.106C} and IXPE's detection of linear X-ray polarization \citep{2023Natur.619...41M} from GCMC both suggest X-ray reflection as the dominant mechanism, with past outbursts from Sgr A$^{\star}$ as a likely source of external X-ray illumination. Low energy cosmic rays could contribute to a baseline \FeKa emission, which would be detectable if a GCMC is not otherwise illuminated by an external X-ray source, such as the case of Sgr B2 after two decades of X-ray luminosity decay \citep{2022icrc.confE.288R}.  While X-ray variability time scale has been previously used to distinguish between the two scenarios for GCMCs, fine spectral features like secondary emission lines from multiply-ionized atoms in the cosmic-ray proton/ion scenario \citep{2020PASJ...72L...7O} serve as a new method to test the two models for GCMC \FeKa emission, not achievable before the era of XRISM.

In this work, we focus on \Resolve detection of \FeKa line complex from  the GCMC G0.11-0.11, located near a radio feature known as the ``radio arc''. G0.11-0.11 is one of several giant molecular clouds in the Sgr A complex \citep{2025A&A...695A..52S,2021A&A...646A..66P}, located $13'$ or a projected distance of $D \approx 30 \:\rm{pc}$ from Sgr~A$^{\star}$. G0.11-0.11 cloud showed strong \FeKa emission since 2000 in XMM-\textit{Newton} observations. Since then, its \FeKa emission has decreased over time until a new X-ray feature within G0.11-0.11 appeared since 2019 \citep{2025A&A...695A..52S}. In Section \ref{sec:ObsAna} we discuss the processing and analysis of \Resolve data, and in Section \ref{sec:Diag} we discuss the physical diagnostics that can be extracted from just the \FeKa complex.  In Section \ref{sec:Conc} we summarize our findings and calculate the external X-ray flux implied by the \FeKa flux from G0.11-0.11. 

\section{Observations and Data Analysis}
\label{sec:ObsAna}

\begin{figure*}[t!]
    \centering
    \includegraphics[width=2\columnwidth]{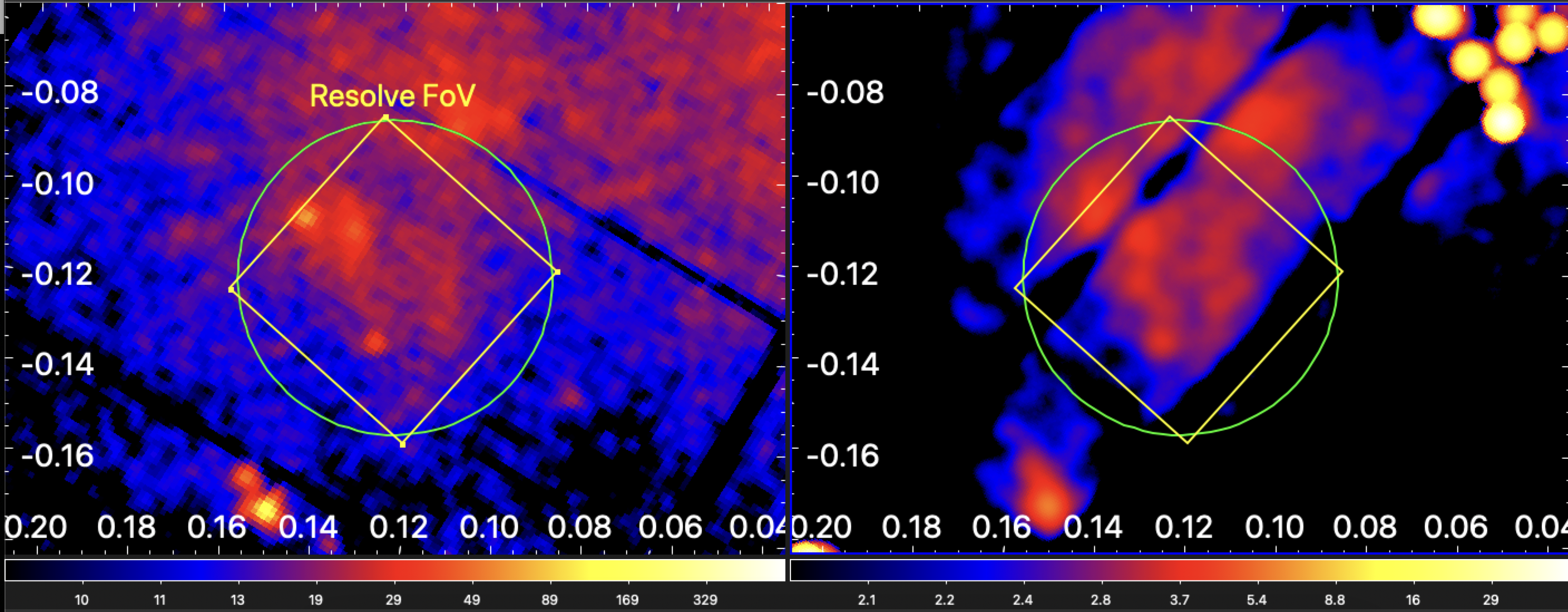}
    \caption{(left) 2025 XMM-\textit{Newton} observation (obs ID $0951870101$) of GCMC G0.11-0.11 between $0.2$ and $12.0 \:~\rm{keV}$ including the regions used to create templates for ray tracing for G0.11-0.11 (green circle, $r=2'$) and the \Resolve field of view (yellow square, 3'x3'). (right) The \textit{XRISM}/Xtend image taken simultaneously with the \Resolve data used in this analysis. In both images, Sgr A$^{\star}$ is just off the upper right corner of the field of view. Both images are in galactic coordinates.}
    \label{fig:XMMtemplates}
\end{figure*}

We download the XRISM observation with ID $201052010$, an $\approx 120 \:\rm{ks}$ exposure with \Resolve targeting G0.11-0.11 starting on 16 March 2025. Using \texttt{heasoft v6.35}, we used \texttt{xapipeline} to reprocess the raw data into cleaned events files. We apply additional screening to remove pixel-pixel coincident events, anomalous low-resolution secondary (Ls) events, and periods of high particle background. We also exclude data from pixel 12, the calibration pixel, and pixel 27, which has unsuitable gain variation compared to the rest of the detector.

Using \texttt{xselect v2.5}, we extract the spectrum of the entire field of view of \Resolve, excluding the pixels noted above. The X-ray emission region in the G0.11-0.11 molecular cloud has roughly the same size as Resolve's field of view and spatial resolution ($\approx 3'$), meaning there is no usable background region in the exposure. We therefore extract the source spectrum from the entire FoV and generate a simulated background using the \texttt{rslnxbgen} command\citep{2021SPIE11444E..5DL}\footnote{Background simulation described in more detail \href{https://heasarc.gsfc.nasa.gov/docs/xrism/analysis/nxb/resolve_nxb_db.html}{here}}.

Subsequently, we use \texttt{rslmkrmf} to generate a large-size RMF; the large-size RMFs include the gaussian cores, exponential tails, silicon instrumental lines, and escape peaks, and produce identical results compared to extra-large RMF for our analysis. We then use \texttt{xaexpmap} to generate an exposure map, taking care to carry over the event and pixel filters discussed above.

Finally, we use \texttt{xaarfgen} to generate an ARF for analysis. Given that G0.11-0.11 is an extended source with irregular shape, we used the image template functionality to simulate ray tracing and generate ARFs, creating an image template from a $4'$-diameter circular slice of the recent XMM-\textit{Newton} observation (ID $0951870101$) centered on G0.11-0.11, which was taken only 12 days after the XRISM observation (the green circle in Figure \ref{fig:XMMtemplates}).  We use this template to generate a ray tracing simulations with $10^5$ source photons, and then create an ARF for all subsequent analysis.

\subsection{Spectral Fitting}

\begin{figure*}
    \centering
    \includegraphics[width=2\columnwidth]{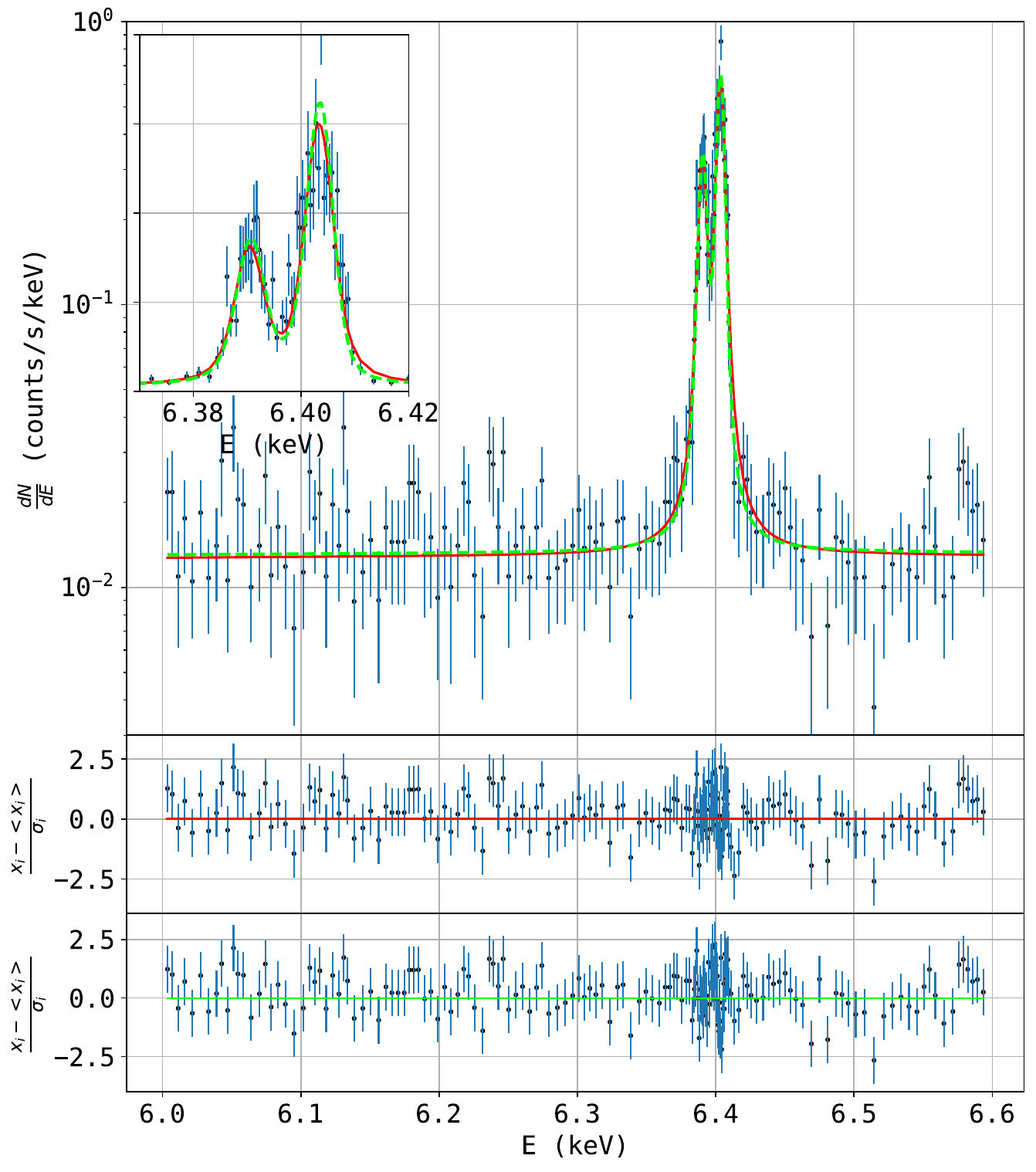}
    \caption{The \Resolve spectrum in $6.0-6.6 \:\rm{keV}$ for G0.11-0.11 (blue), fitted with the two Lorentzian model (red line) and the multiple Lorentzian model (green dotted). The inset shows the \FeKa doublet between 6.36-6.42 keV on a linear scale.}
    \label{fig:RadioSpectrum}
\end{figure*}

Using \texttt{xspec v12.15.0}, we load the spectrum, ARF, RMF, and simulated non-X-ray background, and restrict our analysis to the $6.0 - 6.6 \:~\rm{keV}$ band where the Fe K$\alpha$ line is the only detectable line emission. This range includes energies at which the Compton shoulder (up to $\sim 200 \:\rm{eV}$ below the \FeKa line complex) from X-ray reflection and secondary emission lines from cosmic ray proton/ion ionization (as in \cite{2020PASJ...72L...7O} at energies just above $6.4 \:\rm{keV}$), detectable by \Resolve if present. Figure \ref{fig:RadioSpectrum} shows the extracted spectrum for G0.11-0.11, binned to a minimum signal-to-noise $S/N > 3$ for each bin. 

As shown in Figure \ref{fig:RadioSpectrum}, for the first time the \FeKa emission from GCMC G0.11-0.11 is resolved into distinct \FeKone and \FeKtwo lines. The unprecedented spectral resolution of \Resolve necessitates a Lorentzian model to fit each line, which has broader wings than a Gaussian line. To adequately model the doublets, we develop two parallel spectral models for the \FeKa complex, both including X-ray absorption via \texttt{tbabs} and a power-law continuum.

The first model, ``Two-Lorentzian'', (2L, the red line in Figure \ref{fig:RadioSpectrum}), is a simple phenomenological approach. This model includes two Lorentzian emission lines with peak rest-frame energies at $E_{1}=6.4040 \:\rm{keV}$ and $E_{2}=6.3910 \:\rm{keV}$. The Lorentzian widths and intensities of the two lines are fitted independently, while a single redshift is fitted to both lines simultaneously. In \texttt{xspec}, this model has the form \texttt{tbabs * (powerlaw + zashift * (lorentz + lorentz))}.

The second model (``Multi-Lorentzian'', mL, the lime green dotted line in Figure \ref{fig:RadioSpectrum}) uses the line energy centroids and widths of the constituent Lorentzian components of the \FeKone and \FeKtwo lines from laboratory measurements in \cite{1997PhRvA..56.4554H}, where \FeKone has four Lorentzian subcomponents and \FeKtwo has three. For this model, the relative strengths and widths of each subcomponent are held constant within \FeKone and \FeKtwo, but the total intensities of \FeKone and \FeKtwo are allowed to vary freely, as is the joint redshift. In \texttt{xspec}, this model has the form \texttt{tbabs * (powerlaw + zashift * (constant * (lorentz + lorentz + lorentz + lorentz) + constant *(lorentz + lorentz + lorentz)))}, where the width and magnitude of each Lorentzian sub-component are given in Table 2 of \cite{1997PhRvA..56.4554H}.

To model interstellar absorption, we use the \texttt{tbabs} model with cross-section data from \cite{Verner1996} and abundance from \cite{Wilms2000}. Because the spectrum in the chosen 6.0-6.6 keV energy range does not strongly constrain the absorbing column density $n_H$, we fix $n_H = 5 \times 10^{22}\rm~cm^{-2}$. This value was determined based on a fit to the $2-10 \:\rm{keV}$ spectrum observed simultaneously via XRISM-\textit{XTEND}, which obtained $n_H = (5 \pm 0.5) \times 10^{22} \:\rm{cm^{-2}}$. By fixing $n_H$ to a constant value, we neglect the practically linear scaling in flux normalization that would occur over our narrow energy window if $n_H$ were larger or smaller; the normalizations of the power law continuum and the \FeKa fluxes would change in lockstep by about $\lesssim1\%$ if $n_H$ is changed by $10\%$, so our choice to fix $n_H$ does not impact our analysis results based on relative fluxes.  Over the chosen small energy range, the slope of the power-law continuum cannot be constrained, so we fix it to zero, effectively a constant background with respect to energy. These choices do not change the results reported herein.

Table 1 summarizes the fitting results in both the 2L and mL models with $95\%$ error bars in all cases. In both models, the best-fit normalization of the power-law continuum is $(1.21 \pm 0.07) \times 10^{-4} \:\rm{photons/s/cm^2/keV}$. In the 2L model the ratio of flux in $K\alpha_1$/$K\alpha_2$ is $r = 1.9 \pm 0.3$, within uncertainty to the predicted value from the quantum mechanical definition of the \FeKa transition. In the mL model, we obtain a ratio of line fluxes $1.8 \pm 0.2$. The line FWHM $f$ are not fitted parameters in the mL model, instead being defined by the families of Lorentzian sub-lines used to compose each line. While it might be appropriate to add a convolved Gaussian instrumental width, when we add such a component as a free parameter the width goes to zero. 

The insert in Figure \ref{fig:RadioSpectrum} zooms in on the \FeKa doublet in a linear scale, showing that both the 2L and mL models fit to the spectrum comparably well, with reduced $\chi^2$ of $113.22 /122 = 0.928 $ for 2L and $115.42 / 122 = 0.946$ for mL. 

\begin{table}[]
\centering
\caption{Spectral fitting results for the Two-Lorentzian (2L) and multiple-Lorentzian (mL) models.}
\begin{tabular}{lccc}
\hline \hline
Parameter & Units & 2L & mL \\ \hline
$n_H$ & $10^{22}/\rm{cm}^2$ & $\equiv5$  & $\equiv5$ \\
$\Gamma$ & & $\equiv0$  & $\equiv0$ \\
$N_0$ & $\rm{10^{-5}/s/cm^2/keV}$ & $12.1 \pm 0.7$  & $12.1 \pm 0.7$ \\
$z$ & $\times 10^{-4}$ & $1.0 \pm 0.4$ & $0.3\pm0.4$ \\
\FeKone $N_1$ & $\rm{10^{-5}/s/cm^2}$ & $4.8 \pm 0.5$  & $4.1 \pm 0.3$ \\
\FeKone FWHM & $\rm{eV}$ & $3.2 \pm 0.6$  & fixed \\
\FeKtwo $N_2$ & $\rm{10^{-5}/s/cm^2}$ & $2.5 \pm 0.3$  & $2.3 \pm 0.2$ \\
\FeKtwo FWHM & $\rm{eV}$ & $3.4 \pm 0.7$  & fixed \\
$N_1/N_2$ & & $1.9 \pm 0.3$ & $1.8 \pm 0.2$ \\
$\chi^2/\rm{d.o.f}$ & & $0.928$ & $0.946$ \\ \hline
\end{tabular}
\end{table}

\section{Fe K${\alpha}$ Line Diagnostics}
\label{sec:Diag}

\subsection{Molecular Cloud Velocity with Line Centroids}

The redshifts, tightly constrained by the 2L and mL models, suggest an observational recessional velocity of G0.11-0.11 of $30 \pm 12$ or $9 \pm 12 \:\rm{km/s}$, respectively. Besides the statistical uncertainty from the fitting, there is also a redshift uncertainty imparted by the $\approx 0.3 \:\rm{eV}$ absolute energy scale uncertainty of \Resolve at $6.4 \:\rm{keV}$ \citep{2024SPIE13093E..1PE}. This scale uncertainty corresponds to a $14 \:\rm{km/s}$ velocity uncertainty at the \FeKa lines, making the combined uncertainty of the redshift $\pm 12_{fit} \pm 14_{scale} \:\rm{km/s}$. Before comparisons with radio data can be conducted, two modifications to these measured velocities must be applied. 

First, the observed recessional velocity must be corrected to the solar system barycenter. Using the \verb|radial_velocity_correction| function from \texttt{astropy.coordinates}, we calculate that Earth's motion with respect to the barycenter, projected along the vector towards G0.11-0.11, a velocity correction of $+30 \:\rm{km/s}$ during the observation. Second, because the solar system is moving with respect to the ``Local Standard of Rest'' (LSR, the average motion of stars in the solar neighborhood), our measured radial velocity must be corrected into the LSR frame \citep{2010MNRAS.403.1829S}. Again using \texttt{astropy}'s transformation package, we calculate the LSR correction along the vector to G0.11-0.11 as $+11 \:\rm{km/s}$. Both these adjustments are \textit{towards} the galactic center, so obtaining $v_{\rm{LSR}}$ of G0.11-0.11 requires an adjustment upwards.

Added to our fitted radial velocities, these corrections suggest $v_{\rm{LSR}} = 71 \pm 12_{fit} \pm 14_{scale}$ and $50 \pm 12_{fit} \pm 14_{scale} \:\rm{km/s}$ for 2L and mL, respectively. The tension between the redshifts of the models is likely related to the asymmetry of the \FeKa lines implied by the stacked Lorentzian subcomponents from \cite{1997PhRvA..56.4554H} versus the symmetrical single Lorentzian profiles assumed by the 2L model. Specifically, because the mL model has an asymmetric skew imparted by the stacking of the Lorentzian sub-components, a smaller redshift is required to fit the data than in the 2L model. The errors of each fitted redshift show that \Resolve has the capability to achieve substantial radial velocity precision on very narrow lines.

\Resolve is approaching an energy resolution where the orbital motion of the satellite itself ($\pm 8 \:\rm{km/s}$) could measurably bias the energy centroids of photons to higher or lower values at different times throughout an exposure. Similar to the case discussed in \cite{2025A&A...702A.147X} for NGC 3783, this effect is still small compared to the other components of equivalent width ($\Delta E/E \approx v/c \approx 3 \times 10^{-5}$ giving $\Delta E \approx 0.2 \:\rm{eV}$). Instead of adjusting photon energies at this small scale, we instead note that a small amount of line width is due to the motion of XRISM through its orbit, as this effect is roughly symmetrical. Future works may develop analysis techniques to remote this broadening effect.

\subsection{Line Width}

Our data do not provide strong discrimination between the 2L and mL models. In laboratory results, the quantum mechanical FWHM of the \FeKa lines have been measured \citep{1997PhRvA..56.4554H,PhysRevA.10.2027} as $\rm FWHM = 2.55 \:\rm{eV}$ and $FWHM = 3.14 \:~\rm{eV}$ for \FeKone and \FeKtwo respectively. Our measurements of line FWHM are $FWHM = 3.2 \pm 0.6$ and $FWHM = 3.4 \pm 0.7 \:\rm{eV}$ for \FeKone and \FeKtwo, leaving little room for additional line broadening effect at more than the $\sim\rm{eV}$ level.

Assuming that the instrumental broadening imparted by \Resolve and the physical broadening by any Doppler or thermal effects in G0.11-0.11 are Gaussian in form, and that the profiles of the \FeKa lines are two Lorentzians, the total profile would be a Voight profile. An approximation (Equation 4a in \cite{1977JQSRT..17..233O} and Equation 8 in \cite{1973JOSA...63..987K}) can be used to decompose the FWHM of the constituent Gaussain $f_G$ and Lorentzian $f_L$ components from the overall Voight FWHM $F_V$.

$$ F_V = \frac{1}{2} \left( C_1 F_L + \sqrt{C_2 F_L^2+4 C_3 F_G^2} \right) $$

\noindent with $C_1 = 1 + 0.099 \ln2$, $C_2 = (1 - 0.099 \ln2)^2$, and $C_3 = \ln2$. With the observed $F_V = 3.2 \pm 0.6 \:~\rm{eV}$ observed in Figure \ref{fig:RadioSpectrum} for \FeKone and inherent $F_L = 2.55 \:\rm{eV}$, the remaining Gaussian portion of the width is $F_G = 1.7 \pm 0.6 \:\rm{eV}$. This additional component could be produced by Doppler or thermal broadening from material in G0.11-0.11.

The extraordinary small line width of the \FeKa complex from G0.11-0.11, comparable to the inherent widths of the \FeKa lines and the spectral resolution of \Resolve, allows for \FeKa line diagnostics to serve as precision radial velocity measurement with $R = \frac{E}{\Delta E} > 1000$ when observed with \Resolve.  Correspondence with velocity measurements of GCMC via radio observations is discussed below in Section \ref{sec:Conc}.

\subsection{Cosmic Ray Ionization vs. X-ray Reflection Models}

With unprecedented spectral resolution and sensitivity of \Resolve to line features, a new window opens to search for Compton shoulder and/or secondary line features that can serve as direct and independent tests of the X-ray reflection and cosmic ray ionization models for \FeKa emission in GCMC. 

The Compton shoulder has been sought after as a spectral signature for the X-ray reflection scenario, where \FeKa photons downshifted by $13 - 200 \:\rm{eV} $ after scattered by ambient bound electrons \citep{1998MNRAS.297.1279S} would appear as an excess above the continuum around $6.2 - 6.4 \:\rm{keV}$, with energy and magnitude depending on the geometry of the scattering system. For the G0.11-0.11 cloud, we find no excess in summation by adding the normalized residuals in the 2L and mL models in the energy range $6.2 - 6.38 \:\rm{keV}$. By adding a box-like component to the mL model, we can establish a $3\sigma$ upper limit on the total flux in the Compton shoulder for G0.11-0.11. We find the upper limit photon flux to be $<0.014\:\rm{photons/s}$ in $6.2 - 6.38 \:\rm{keV}$, or $< 1.7\%$ of the primary line flux.

While detection of the Compton shoulder would serve as a firm confirmation of the X-ray reflection scenario, the lack of it could be attributed to several reasons under the same scenario. First, it might not have been long enough for multi-scattering since the arrival of the X-ray illumination wavefront.  As demonstrated in \cite{1998MNRAS.297.1279S} and \cite{2011ApJ...740..103O}, the relative strength of the Compton shoulder compared to the primary peaks increases with time as \FeKa photons are scattered. Figure 8 of \cite{2011ApJ...740..103O} in particular shows how the shoulder-to-peak flux ratio increases to $\approx 0.1$ on timescales of $t \sim 3 R/c$ for $R$ the radius of the molecular cloud, depending on geometry. With G0.11-0.11 having angular size $\approx 2'$ or a physical size at $8 \:\rm{kpc}$ of $\approx 15 \:\rm{ly}$, it could take another several decades before the Compton shoulder reached 10\% of the \FeKa flux. Secondly and perhaps more likely, G0.11-0.11 could be optically thin cloud with an optical depth $\tau \le 0.1$, in which case a \FeKa photon produced therein is unlikely to interact with an ambient iron atom within the cloud. 

%%% SZ: We know from variability study that G0.11 was illuminated around or before 2000, then again arounf 2019.

A spectral signature of cosmic ray proton/ion ionization model is the presence of secondary \FeKa lines after a single cosmic ray proton or heavier ion ejects multiple electrons from an Fe atom. In \cite{2020PASJ...72L...7O}, the authors developed a detailed model for how cosmic ray ions generate secondary emission lines that would be detectable by \Resolve, especially Fe K$\alpha_1$L1 and Fe K$\alpha_2$L1 at $\approx6.420$ and $\approx 6.435 \:\rm{keV}$ respectively. Adopting the model in Figure 4 of \cite{2020PASJ...72L...7O} for a solar-abundance cosmic ray flux with $\alpha = 1$ in the energy range $0.5 - 1000 \:\rm{MeV}$, we create a \texttt{xspec} model with additional Lorentzian profiles assuming that 100\% of the \FeKa flux in the primary peaks is due to cosmic ray ionization.
 
\begin{figure}
    \centering
    \includegraphics[width=\columnwidth]{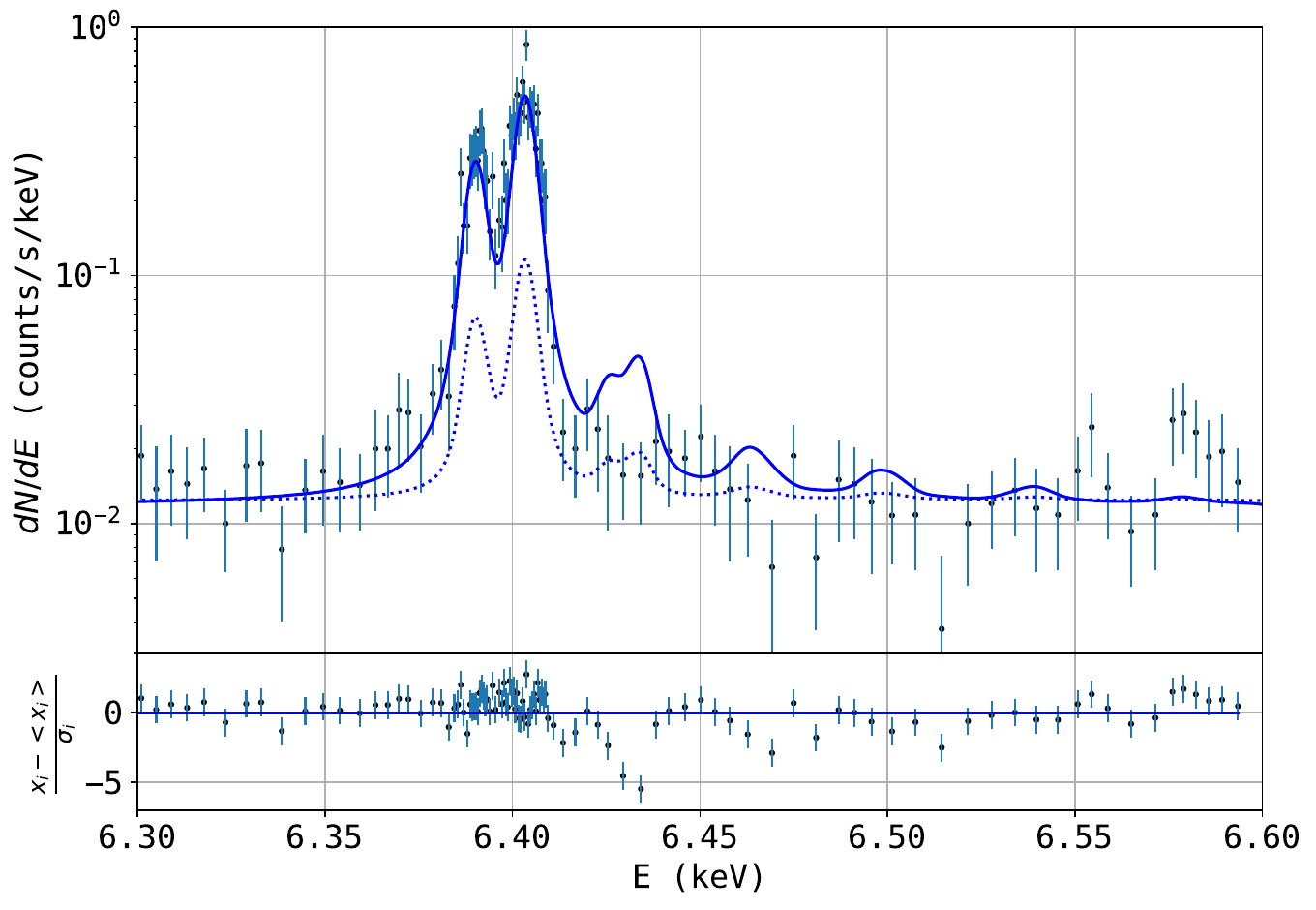}
    \caption{The $\alpha = 1$ CR ionization model (blue solid line) from \cite{2020PASJ...72L...7O} assuming 100\% of \FeKa primary flux from ionization by CR proton and heavier elements, versus the $3\sigma$ upper limit of $20\%$ of the observed \FeKa flux (faint blue dotted).}
    \label{fig:CRCompare}
\end{figure}

As shown in Figure \ref{fig:CRCompare}, the cosmic ray proton/ion model following \cite{2020PASJ...72L...7O} predicts substantial flux at secondary lines approximately $30 \:\rm{eV}$ above \FeKone. The \Resolve data shows no such features; the data exclude a model where 100\% of the \FeKa flux is produced by CR excitation at the $10\sigma$ level. This finding therefore disfavors cosmic ray excitation as the dominant mechanism. Conducting a joint fit with both a mL component representing external X-ray reflection and the CR ionization model sends the amplitude of the CR component to zero due to the total lack of secondary lines. Scaling down the CR ionization model in Figure \ref{fig:CRCompare} obtains a $3\sigma$ upper limit of at most 20\% of the \FeKa emission from cosmic ray proton/ion ionization.

%external X-ray ionization as the progenitor of the \FeKa flux observed from G0.11-0.11, an independent piece of evidence along with detection of linear X-ray polarization \citep{2023Natur.619...41M} and X-ray variability \citep{2025A&A...695A..52S} measurements, both favoring the X-ray reflection model. 

% cannot rule out cosmic ray electron here. So removing the above conclusion.

\section{Summary and Conclusions}
\label{sec:Conc}

Using new observations at the GCMC G0.11-0.11 with \Resolve we characterize the \FeKa doublet with unprecedented energy resolution. This precise spectroscopy allows for detailed spectral and physical diagnostics of G0.11-0.11 and provides a new method to distinguish between competing scenarios leading to \FeKa emission in GCMC by resolving or constraining expected spectral features.

We measure the FWHM of the \FeKone and \FeKtwo lines at $3.2 \pm 0.6$ and $3.4 \pm 0.7 \:\rm{eV}$ in a two Lorentzian model. Between the instrumental resolution and the inherent width of the \FeKa lines, there is little room for further broadening effects such as thermal or collisional broadening, suggesting that G0.11-0.11 is quite a thermally cool cloud, in agreement with measurements of thermal broadening observed with radio data \citep{2025ApJ...984..157B}. The \FeKa line broadening fit in the 2L model can be well accounted for by the quantum mechanical width of the \FeKa lines plus a small Gaussian component of magnitude $\lesssim 1.7 \pm 0.6 \:\rm{eV}$. This tight constraint on Doppler effect induced line broadening limits the G0.11-0.11 dispersive velocity to $\Delta v \lesssim 70 \:\rm{km/s}$, in agreement with values determined via radio observations in \cite{2025ApJ...984..157B}.

We also measure the radial velocity of G0.11-0.11 using the \FeKa lines. In the mL model, we obtain $v_{\rm{LSR}} = 50 \pm 12_{fit} \pm 14_{scale} \:\rm{km/s}$, slightly different from the redshift measured in the 2L model due to asymmetric line shape of the stacked Lorentzian sub-components. The $v_{\rm{LSR}}$ obtained from the mL model is in good agreement with the velocities reported in \cite{2025ApJ...984..157B} for G0.11-0.11 (see Table 2 for object ID 35) with $v_{\rm{LRS}} = 52 \:\rm{km/s}$ via measurements of molecular lines from $\rm{HCNO}$, $\rm{HCN}$, and $\rm{HC_3N}$. Alternatively, G0.11-0.11 is identified with a lower radial velocity of $25 - 45 \:\rm{km/s}$ in \cite{2011PASJ...63..763T} and \cite{2025A&A...695A..52S}, suggesting that perhaps only a portion of the overall radio cloud is contributing X-rays. Future observation of changing \FeKa emission from G0.11-0.11 will constrain these scenarios in more detail. With our mL estimate for the radial velocity of G0.11-0.11 being within uncertainties to radio measurements, our \Resolve data supports the production of \FeKa X-rays in the same clouds identified in radio observations of the GC \citep{2025ApJ...984..157B}.

Besides the sharp \FeKa doublet, the $6.0-6.6 \:\rm{keV}$ spectrum of G0.11-0.11 has no secondary features. The absence of secondary lines from cosmic ray proton/ion ionization of multiple electrons as predicted in \citep{2020PASJ...72L...7O} suggests that cosmic ray proton/ion ionization is not a significant contributor to the \FeKa flux in this case; indeed, Figure \ref{fig:CRCompare} shows how the models developed in \cite{2020PASJ...72L...7O} constrain the contribution of CR ionization to less than $20\%$ of the observed \FeKa flux given the absence of the secondary Fe K$\alpha_1$L1 and Fe K$\alpha_2$L1 lines about $\sim 30 \:\rm{eV}$ above the primary \FeKa doublet.

Finally, we see no evidence for a Compton shoulder, which would appear roughly $13 - 200 \:\rm{eV}$ below the primary \FeKa doublet \citep{1998MNRAS.297.1279S}.  The lack of this shoulder has several possible physical explanations, including that insufficient time has elapsed since G0.11-0.11's illumination by an external X-ray source for \FeKa photons to be reprocessed into Compton scattered photons at lower energies. More importantly, G0.11-0.11 could be optically thin enough such that locally-produced \FeKa photons escape without Compton scatting to lower energies. Further observations to monitor the spectral and flux variability of G0.11-0.11 over the next few years will test this hypothesis.

\subsection{X-ray Reflection Model: Constraining past Sgr~A$^{\star}$ luminosity}

With a precise measurement of the total \FeKa flux from G0.11-0.11, we can use the geometrical X-ray reflection model in \cite{1998MNRAS.297.1279S} and \cite{2015ApJ...815..132Z} to estimate the luminosity of Sgr~A$^{\star}$ as the external illuminating source, comparing with values obtained in other works \citep{2013A&A...558A..32C,2015ApJ...815..132Z} by using similar assumptions about the GC environment. Equation 2 in \cite{1998MNRAS.297.1279S} relates $F_{6.4}$, the photon flux in the \FeKa lines in $\rm{photons/s/cm^2}$, to $I(\rm{8 \: keV})$ the X-ray intensity of the external illuminating source at $8~\:\rm{keV}$.

\begin{equation}
    \frac{F_{6.4}}{(\rm{/s/cm^2})} = \phi \frac{\Omega}{4 \pi D_{cm}^2} \tau_T Z_{Fe} I(\rm{8 \: keV)}
\end{equation}

\noindent where $\phi \approx 1.13$ is a factor accounting for an assumed power law incident spectrum with $\Gamma=2$, $\Omega$ is the solid angle covered by the cloud from the perspective of the illuminating source in steradians, $D_{\rm{cm}}$ is the distance from Earth to the cloud in centimeters, $\tau_T$ is the optical depth of the cloud to \FeKa photons, and $Z_{Fe}$ is the abundance of iron in the cloud relative to solar abundance ($3.3 \times 10^{-5}$). This approach to calculating  contains substantial uncertainties on cloud density, iron abundance, shape, size, and illumination fraction, enabling comparison with previous estimates by following similar assumptions.

Via the approach in \cite{1998MNRAS.297.1279S}, $I(\rm{8 \: keV})$ can be expressed in terms of $L_8 = I(\rm{8 \: keV}) \times 8^2 \times 1.6\times 10^{-9} \:\rm{erg/s}$, the X-ray luminosity in an $8\:\rm{keV}$-wide band centered at $8\:\rm{keV}$. Equation 2 from \cite{2015ApJ...815..132Z} can be directly adapted from the case of Sgr B2 (which has an angular size of $1.5'$) to G0.11-0.11 by making similar assumptions of iron abundance and illumination fraction but modifying for G0.11-0.11's angular size of $1'$  (decreasing the prefactor of Equation 2 by $(\frac{1'}{1.5'})^2 = 0.44$). We can express required the illuminating source luminosity as:

\begin{equation}
    \frac{L_8}{(\rm{erg/s})} = \frac{1.3 \times 10^{39}}{Z_{Fe}} \left(\frac{F_{6.4}}{10^{-4}}\right) \left(\frac{0.1}{\tau_T}\right) \left(\frac{d}{100 \:\rm{pc}}\right)^{2}
\end{equation}

\noindent where $d$ is the distance from the illuminating source, in this case Sgr~A$^{\star}$, to G0.11-0.11. The distance between G0.11-0.11 and Sgr~A$^{\star}$ is poorly constrained and is a dominant contributor to the uncertainty of illuminating luminosity, along with degeneracies introduced by abundance, angular size, and illuminated fraction. We take the measured total column density $N_H=5\times10^{22}\rm~cm^{-2}$ as the upper limit for the column density of G0.11-0.11, giving a corresponding upper limit on optical depth of $\tau_{T} \approx 0.06$. In this case, we use $\tau_T \approx 0.1$. For abundance we adopt $Z_{Fe} = 1.3$ as measured in \cite{2010ApJ...719..143T}.

%%The minimum value for $d$ is the projected distance between Sgr~A$^{\star}$ and G0.11-0.11 in the plane of the sky, $d>30 \:\rm{pc}$, but some estimates \citep{2025A&A...695A..52S,2023Natur.619...41M} suggest that G0.11-0.11 may have be positioned about 20 pc behind Sgr~A$^{\star}$ with respects to the plane of the sky, so that $d \approx 34 \:\rm{pc}$.

The total \FeKa flux is $F_{6.4}=N_1 + N_2 =(7.3\pm0.6)\times10^{-5}\rm~ph/s/cm^{2}$, giving a required illuminating luminosity of $L_8 \approx 10^{39} \left(\frac{d}{100 \:\rm{pc}}\right)^{2} \:\rm{erg/s}$. With the most recently estimated distance between G0.11-0.11 and Sgr~A$^{\star}$ as $d \approx 34$~pc \citep{2025A&A...695A..52S}, the required luminosity would be $L_8 \approx 10^{38}~\rm{erg/s}$. This is several orders of magnitudes higher than the current X-ray luminosity of Sgr~A$^{\star}$ in a quiescent state, but is comparable to the historic X-ray luminosity estimated by other works \citep{2025A&A...695A..52S,2015ApJ...815..132Z}, showing that our \Resolve measurements enable comparable studies of the history of Sgr~A$^{\star}$.

Exactly when such significant X-ray outburst happened depends on the exact location of each individual cloud and whether one or multiple outbursts happened in the past a few hundred years. IXPE measurements \citep{2023Natur.619...41M} suggest that a major Sgr~A$^{\star}$ outburst at such level happened about 200 years ago, and several previous works  \citep{2013A&A...558A..32C,2018A&A...610A..34C,2025A&A...695A..52S} have argued that a ``two flare" model is a feasible alternative. In the ``two flare'' case, G0.11-0.11 and other GCMCs are illuminated by a $\sim 230$ year old flare while the Bridge cloud is illuminated by a $\sim 130$ year old flare. The more recent flare would eventually travel to G0.11-0.11 and illuminate it again in a few decades. Future X-ray monitoring of GCMCs will test the ``one-flare" and ``two-flare" models, characterize the unique dynamics of each GCMC, and create a more comprehensive map of the Galactic center region.

\software{FTools \citep{FTools}, Xspec \citep{Xspec}, DS9 \citep{DS9}}

\acknowledgments

We gratefully acknowledge the scientific guidance and insight of Dr. Kaya Mori and Dr. Vincent Tatischeff, who substantially contributed to the interpretation of our results. Unending gratitude is also due to Dr. Koji Mukai and Dr. Ayşegül Tümer for technical advice vital to the processing of the \Resolve data used herein. This work is supported by NASA XRISM cycle-1 guest observer grant number 80NSSC25K7844. No AI or LLM tools were used in the construction of this manuscript or the conduct of the science herein. We thank the anonymous referee for their comments and guidance, which have substantially improved this work.

\bibliography{main}{}

\begin{thebibliography}{}
\expandafter\ifx\csname natexlab\endcsname\relax\def\natexlab#1{#1}\fi
\providecommand{\url}[1]{\href{#1}{#1}}
\providecommand{\dodoi}[1]{doi:~\href{http://doi.org/#1}{\nolinkurl{#1}}}
\providecommand{\doeprint}[1]{\href{http://ascl.net/#1}{\nolinkurl{http://ascl.net/#1}}}
\providecommand{\doarXiv}[1]{\href{https://arxiv.org/abs/#1}{\nolinkurl{https://arxiv.org/abs/#1}}}

\bibitem[{{Arnaud}(1996)}]{Xspec}
{Arnaud}, K.~A. 1996, in Astronomical Society of the Pacific Conference Series, Vol. 101, Astronomical Data Analysis Software and Systems V, ed. G.~H. {Jacoby} \& J.~{Barnes}, 17

\bibitem[{{Battersby} {et~al.}(2025){Battersby}, {Walker}, {Barnes}, {Ginsburg}, {Lipman}, {Alboslani}, {Hatchfield}, {Bally}, {Glover}, {Henshaw}, {Immer}, {Klessen}, {Longmore}, {Mills}, {Molinari}, {Smith}, {Sormani}, {Tress}, \& {Zhang}}]{2025ApJ...984..157B}
{Battersby}, C., {Walker}, D.~L., {Barnes}, A., {et~al.} 2025, \apj, 984, 157, \dodoi{10.3847/1538-4357/adb844}

\bibitem[{{Capelli} {et~al.}(2011){Capelli}, {Warwick}, {Porquet}, {Gillessen}, \& {Predehl}}]{2011A&A...530A..38C}
{Capelli}, R., {Warwick}, R.~S., {Porquet}, D., {Gillessen}, S., \& {Predehl}, P. 2011, \aap, 530, A38, \dodoi{10.1051/0004-6361/201116574}

\bibitem[{{Chuard} {et~al.}(2018){Chuard}, {Terrier}, {Goldwurm}, {Clavel}, {Soldi}, {Morris}, {Ponti}, {Walls}, \& {Chernyakova}}]{2018A&A...610A..34C}
{Chuard}, D., {Terrier}, R., {Goldwurm}, A., {et~al.} 2018, \aap, 610, A34, \dodoi{10.1051/0004-6361/201731864}

\bibitem[{{Clavel} {et~al.}(2012){Clavel}, {Terrier}, {Goldwurm}, {Morris}, {Ponti}, {Soldi}, \& {TRAP}}]{2012int..workE.106C}
{Clavel}, M., {Terrier}, R., {Goldwurm}, A., {et~al.} 2012, in Proceedings of ``An INTEGRAL view of the high-energy sky (the first 10 years)'' - 9th INTEGRAL Workshop and celebration of the 10th anniversary of the launch (INTEGRAL 2012). 15-19 October 2012. Bibliotheque Nationale de France, 106, \dodoi{10.22323/1.176.0106}

\bibitem[{{Clavel} {et~al.}(2013){Clavel}, {Terrier}, {Goldwurm}, {Morris}, {Ponti}, {Soldi}, \& {Trap}}]{2013A&A...558A..32C}
{Clavel}, M., {Terrier}, R., {Goldwurm}, A., {et~al.} 2013, \aap, 558, A32, \dodoi{10.1051/0004-6361/201321667}

\bibitem[{{Dogiel} {et~al.}(2009){Dogiel}, {Cheng}, {Chernyshov}, {Bamba}, {Ichimura}, {Inoue}, {Ko}, {Kokubun}, {Maeda}, {Mitsuda}, \& {Yamasaki}}]{2009PASJ...61..901D}
{Dogiel}, V., {Cheng}, K.-S., {Chernyshov}, D., {et~al.} 2009, \pasj, 61, 901, \dodoi{10.1093/pasj/61.4.901}

\bibitem[{{Eckart} {et~al.}(2024){Eckart}, {Brown}, {Chiao}, {Cumbee}, {Fujimoto}, {Hell}, {Hoshino}, {Ishisaki}, {Kelley}, {Kenyon}, {Kilbourne}, {Kitamoto}, {Leutenegger}, {Lockard}, {Loewenstein}, {Magee}, {Mizumoto}, {Porter}, {Sato}, {Sawada}, {Shah}, {Shipman}, {Sneiderman}, {Takei}, {Tsujimoto}, {de Vries}, {Watanabe}, {Witthoeft}, {Wolfs}, {Yamada}, \& {Yaqoob}}]{2024SPIE13093E..1PE}
{Eckart}, M.~E., {Brown}, G.~V., {Chiao}, M.~P., {et~al.} 2024, in Society of Photo-Optical Instrumentation Engineers (SPIE) Conference Series, Vol. 13093, Space Telescopes and Instrumentation 2024: Ultraviolet to Gamma Ray, ed. J.-W.~A. {den Herder}, S.~{Nikzad}, \& K.~{Nakazawa}, 130931P, \dodoi{10.1117/12.3019276}

\bibitem[{Heasarc(2014)}]{FTools}
Heasarc. 2014, {HEAsoft: Unified Release of FTOOLS and XANADU}, Astrophysics Source Code Library, record ascl:1408.004

\bibitem[{{H{\"o}lzer} {et~al.}(1997){H{\"o}lzer}, {Fritsch}, {Deutsch}, {H{\"a}rtwig}, \& {F{\"o}rster}}]{1997PhRvA..56.4554H}
{H{\"o}lzer}, G., {Fritsch}, M., {Deutsch}, M., {H{\"a}rtwig}, J., \& {F{\"o}rster}, E. 1997, \pra, 56, 4554, \dodoi{10.1103/PhysRevA.56.4554}

\bibitem[{{Joye} \& {Mandel}(2003)}]{DS9}
{Joye}, W.~A., \& {Mandel}, E. 2003, in Astronomical Society of the Pacific Conference Series, Vol. 295, Astronomical Data Analysis Software and Systems XII, ed. H.~E. {Payne}, R.~I. {Jedrzejewski}, \& R.~N. {Hook}, 489

\bibitem[{{Kielkopf}(1973)}]{1973JOSA...63..987K}
{Kielkopf}, J.~F. 1973, Journal of the Optical Society of America (1917-1983), 63, 987

\bibitem[{Lee \& Salem(1974)}]{PhysRevA.10.2027}
Lee, P.~L., \& Salem, S.~I. 1974, Phys. Rev. A, 10, 2027, \dodoi{10.1103/PhysRevA.10.2027}

\bibitem[{{Loewenstein} {et~al.}(2021){Loewenstein}, {Hill}, {Holland}, {Miller}, {Yaqoob}, {Doyle}, {Hall}, {Braun}, {Baluta}, {Mukai}, {Terada}, {Tashiro}, {Takahashi}, {Nobukawa}, {Mizuno}, {Tamura}, {Uno}, {Watanabe}, {Ebisawa}, {Eguchi}, {Fukazawa}, {Iizuka}, {Katsuda}, {Kitaguchi}, {Kubota}, {Nakashima}, {Nakazawa}, {Odaka}, {Ohno}, {Ota}, {Sato}, {Sugawara}, {Shidatsu}, {Tamba}, {Terashima}, {Tsuboi}, {Uchida}, {Uchiyama}, \& {Yamauchi}}]{2021SPIE11444E..5DL}
{Loewenstein}, M., {Hill}, R.~S., {Holland}, M.~P., {et~al.} 2021, in Society of Photo-Optical Instrumentation Engineers (SPIE) Conference Series, Vol. 11444, Society of Photo-Optical Instrumentation Engineers (SPIE) Conference Series, ed. J.-W.~A. {den Herder}, S.~{Nikzad}, \& K.~{Nakazawa}, 114445D, \dodoi{10.1117/12.2560840}

\bibitem[{{Marin} {et~al.}(2023){Marin}, {Churazov}, {Khabibullin}, {Ferrazzoli}, {Di Gesu}, {Barnouin}, {Di Marco}, {Middei}, {Vikhlinin}, {Costa}, {Soffitta}, {Muleri}, {Sunyaev}, {Forman}, {Kraft}, {Bianchi}, {Donnarumma}, {Petrucci}, {Enoto}, {Agudo}, {Antonelli}, {Bachetti}, {Baldini}, {Baumgartner}, {Bellazzini}, {Bongiorno}, {Bonino}, {Brez}, {Bucciantini}, {Capitanio}, {Castellano}, {Cavazzuti}, {Chen}, {Ciprini}, {De Rosa}, {Del Monte}, {Di Lalla}, {Doroshenko}, {Dov{\v{c}}iak}, {Ehlert}, {Evangelista}, {Fabiani}, {Garcia}, {Gunji}, {Hayashida}, {Heyl}, {Ingram}, {Iwakiri}, {Jorstad}, {Kaaret}, {Karas}, {Kitaguchi}, {Kolodziejczak}, {Krawczynski}, {La Monaca}, {Latronico}, {Liodakis}, {Maldera}, {Manfreda}, {Marinucci}, {Marscher}, {Marshall}, {Massaro}, {Matt}, {Mitsuishi}, {Mizuno}, {Negro}, {Ng}, {O'Dell}, {Omodei}, {Oppedisano}, {Papitto}, {Pavlov}, {Peirson}, {Perri}, {Pesce-Rollins}, {Pilia}, {Possenti}, {Poutanen}, {Puccetti}, {Ramsey}, {Rankin}, {Ratheesh}, {Roberts}, {Romani}, {Sgr{\`o}},
  {Slane}, {Spandre}, {Swartz}, {Tamagawa}, {Tavecchio}, {Taverna}, {Tawara}, {Tennant}, {Thomas}, {Tombesi}, {Trois}, {Tsygankov}, {Turolla}, {Vink}, {Weisskopf}, {Wu}, {Xie}, \& {Zane}}]{2023Natur.619...41M}
{Marin}, F., {Churazov}, E., {Khabibullin}, I., {et~al.} 2023, \nat, 619, 41, \dodoi{10.1038/s41586-023-06064-x}

\bibitem[{{Odaka} {et~al.}(2011){Odaka}, {Aharonian}, {Watanabe}, {Tanaka}, {Khangulyan}, \& {Takahashi}}]{2011ApJ...740..103O}
{Odaka}, H., {Aharonian}, F., {Watanabe}, S., {et~al.} 2011, \apj, 740, 103, \dodoi{10.1088/0004-637X/740/2/103}

\bibitem[{{Okon} {et~al.}(2020){Okon}, {Imai}, {Tanaka}, {Uchida}, \& {Tsuru}}]{2020PASJ...72L...7O}
{Okon}, H., {Imai}, M., {Tanaka}, T., {Uchida}, H., \& {Tsuru}, T.~G. 2020, \pasj, 72, L7, \dodoi{10.1093/pasj/psaa055}

\bibitem[{{Olivero}(1977)}]{1977JQSRT..17..233O}
{Olivero}, J. 1977, \jqsrt, 17, 233, \dodoi{10.1016/0022-4073(77)90161-3}

\bibitem[{{Ponti} {et~al.}(2021){Ponti}, {Morris}, {Churazov}, {Heywood}, \& {Fender}}]{2021A&A...646A..66P}
{Ponti}, G., {Morris}, M.~R., {Churazov}, E., {Heywood}, I., \& {Fender}, R.~P. 2021, \aap, 646, A66, \dodoi{10.1051/0004-6361/202039636}

\bibitem[{{Ponti} {et~al.}(2010){Ponti}, {Terrier}, {Goldwurm}, {Belanger}, \& {Trap}}]{2010ApJ...714..732P}
{Ponti}, G., {Terrier}, R., {Goldwurm}, A., {Belanger}, G., \& {Trap}, G. 2010, \apj, 714, 732, \dodoi{10.1088/0004-637X/714/1/732}

\bibitem[{{Rogers} {et~al.}(2022){Rogers}, {Zhang}, {Perez}, {Clavel}, \& {Taylor}}]{2022icrc.confE.288R}
{Rogers}, F., {Zhang}, S., {Perez}, K., {Clavel}, M., \& {Taylor}, A. 2022, in 37th International Cosmic Ray Conference, 288, \dodoi{10.22323/1.395.0288}

\bibitem[{{Sch{\"o}nrich} {et~al.}(2010){Sch{\"o}nrich}, {Binney}, \& {Dehnen}}]{2010MNRAS.403.1829S}
{Sch{\"o}nrich}, R., {Binney}, J., \& {Dehnen}, W. 2010, \mnras, 403, 1829, \dodoi{10.1111/j.1365-2966.2010.16253.x}

\bibitem[{{Stel} {et~al.}(2025){Stel}, {Ponti}, {Haardt}, \& {Sormani}}]{2025A&A...695A..52S}
{Stel}, G., {Ponti}, G., {Haardt}, F., \& {Sormani}, M. 2025, \aap, 695, A52, \dodoi{10.1051/0004-6361/202451359}

\bibitem[{{Sunyaev} \& {Churazov}(1998)}]{1998MNRAS.297.1279S}
{Sunyaev}, R., \& {Churazov}, E. 1998, \mnras, 297, 1279, \dodoi{10.1046/j.1365-8711.1998.01684.x}

\bibitem[{{Tatischeff} {et~al.}(2012){Tatischeff}, {Decourchelle}, \& {Maurin}}]{2012A&A...546A..88T}
{Tatischeff}, V., {Decourchelle}, A., \& {Maurin}, G. 2012, \aap, 546, A88, \dodoi{10.1051/0004-6361/201219016}

\bibitem[{{Terrier} {et~al.}(2010){Terrier}, {Ponti}, {B{\'e}langer}, {Decourchelle}, {Tatischeff}, {Goldwurm}, {Trap}, {Morris}, \& {Warwick}}]{2010ApJ...719..143T}
{Terrier}, R., {Ponti}, G., {B{\'e}langer}, G., {et~al.} 2010, \apj, 719, 143, \dodoi{10.1088/0004-637X/719/1/143}

\bibitem[{{Tsuboi} {et~al.}(2011){Tsuboi}, {Tadaki}, {Miyazaki}, \& {Handa}}]{2011PASJ...63..763T}
{Tsuboi}, M., {Tadaki}, K.-i., {Miyazaki}, A., \& {Handa}, T. 2011, \pasj, 63, 763, \dodoi{10.1093/pasj/63.4.763}

\bibitem[{{Verner} {et~al.}(1996){Verner}, {Ferland}, {Korista}, \& {Yakovlev}}]{Verner1996}
{Verner}, D.~A., {Ferland}, G.~J., {Korista}, K.~T., \& {Yakovlev}, D.~G. 1996, \apj, 465, 487, \dodoi{10.1086/177435}

\bibitem[{{Wilms} {et~al.}(2000){Wilms}, {Allen}, \& {McCray}}]{Wilms2000}
{Wilms}, J., {Allen}, A., \& {McCray}, R. 2000, \apj, 542, 914, \dodoi{10.1086/317016}

\bibitem[{{Xrism Collaboration} {et~al.}(2025){Xrism Collaboration}, {Audard}, {Awaki}, {Ballhausen}, {Bamba}, {Behar}, {Boissay-Malaquin}, {Brenneman}, {Brown}, {Corrales}, {Costantini}, {Cumbee}, {D{\'\i}az Trigo}, {Done}, {Dotani}, {Ebisawa}, {Eckart}, {Eckert}, {Eguchi}, {Enoto}, {Ezoe}, {Foster}, {Fujimoto}, {Fujita}, {Fukazawa}, {Fukushima}, {Furuzawa}, {Gallo}, {Garc{\'\i}a}, {Gu}, {Guainazzi}, {Hagino}, {Hamaguchi}, {Hatsukade}, {Hayashi}, {Hayashi}, {Hell}, {Hodges-Kluck}, {Hornschemeier}, {Ichinohe}, {Ishi}, {Ishida}, {Ishikawa}, {Ishisaki}, {Kaastra}, {Kallman}, {Kanemaru}, {Kara}, {Katsuda}, {Kelley}, {Kilbourne}, {Kitamoto}, {Kobayashi}, {Kohmura}, {Kubota}, {Leutenegger}, {Loewenstein}, {Maeda}, {Markevitch}, {Matsumoto}, {Matsushita}, {McCammon}, {McNamara}, {Mernier}, {Miller}, {Miller}, {Mitsuishi}, {Mizumoto}, {Mizuno}, {Mori}, {Mukai}, {Murakami}, {Mushotzky}, {Nakajima}, {Nakazawa}, {Ness}, {Nobukawa}, {Nobukawa}, {Noda}, {Odaka}, {Ogawa}, {Ogorzalek}, {Okajima}, {Ota}, {Paltani}, {Petre},
  {Plucinsky}, {Porter}, {Pottschmidt}, {Sato}, {Sato}, {Sawada}, {Seta}, {Shidatsu}, {Simionescu}, {Smith}, {Suzuki}, {Szymkowiak}, {Takahashi}, {Takeo}, {Tamagawa}, {Tamura}, {Tanaka}, {Tanimoto}, {Tashiro}, {Terada}, {Terashima}, {Tsuboi}, {Tsujimoto}, {Tsunemi}, {Tsuru}, {T{\"u}mer}, {Uchida}, {Uchida}, {Uchida}, {Uchiyama}, {Ueda}, {Uno}, {Vink}, {Watanabe}, {Williams}, {Yamada}, {Yamada}, {Yamaguchi}, {Yamaoka}, {Yamasaki}, {Yamauchi}, {Yamauchi}, {Yaqoob}, {Yoneyama}, {Yoshida}, {Yukita}, {Zhuravleva}, {Diez}, {Fukumura}, {Li}, {Mehdipour}, {Panagiotou}, {Signorini}, \& {Zhao}}]{2025A&A...702A.147X}
{Xrism Collaboration}, {Audard}, M., {Awaki}, H., {et~al.} 2025, \aap, 702, A147, \dodoi{10.1051/0004-6361/202556000}

\bibitem[{{Zhang} {et~al.}(2015){Zhang}, {Hailey}, {Mori}, {Clavel}, {Terrier}, {Ponti}, {Goldwurm}, {Bauer}, {Boggs}, {Christensen}, {Craig}, {Harrison}, {Hong}, {Nynka}, {Soldi}, {Stern}, {Tomsick}, \& {Zhang}}]{2015ApJ...815..132Z}
{Zhang}, S., {Hailey}, C.~J., {Mori}, K., {et~al.} 2015, \apj, 815, 132, \dodoi{10.1088/0004-637X/815/2/132}

\end{thebibliography}

\end{document}